\documentclass[10pt,twocolumn,showpacs,preprintnumbers,amsmath,amssymb,aps,prb,
superscriptaddress,floatfix,footinbib]{revtex4-2}
\usepackage{graphicx,color}
 \usepackage{dcolumn}
 \usepackage{bm}
 \usepackage{epstopdf}
 \usepackage{tabularx}

\begin{document}


\title{Electric polarization driven by non-collinear spin alignment investigated by first principles calculations}

\author{Sergiy Mankovsky$^*$}
\affiliation{Institute  of  Materials  Science, TU Darmstadt, 64287
  Darmstadt, Germany} 
\affiliation{Department of Chemistry/Phys.\ Chemistry, LMU Munich,
  Butenandtstrasse 11, D-81377 Munich, Germany}
\email{mankovsky@tmm.tu-darmstadt.de}

\author{Svitlana Polesya}
\affiliation{Department of Chemistry/Phys.\ Chemistry, LMU Munich,
  Butenandtstrasse 11, D-81377 Munich, Germany}

\author{Jan Minar}
\affiliation{New Technologies Research Center, University of West
  Bohemia, 30100 Pilsen, Czech Republic}

\author{Hongbin Zhang}
\affiliation{Institute  of  Materials  Science, TU Darmstadt, 64287
  Darmstadt, Germany} 

\author{Hubert Ebert}
\affiliation{Department of Chemistry/Phys.\ Chemistry, LMU Munich,
Butenandtstrasse 11, D-81377 Munich, Germany}

\pacs{}

\date{\today}

\begin{abstract}
We present an approach for first principles investigations on the spin
driven electric polarization in type II multiferroics. We propose a
  parametrization of the polarization with the parameters calculated
  using the Korringa-Kohn-Rostoker Green function (KKR-GF) formalism.
Within this approach the induced electric polarization of a unit cell is
represented in terms of three-site parameters. Those antisymmetric
with respect to spin permutation are seen as an {\it ab initio} based
counter-part to the phenomenological parameters used within the
inverse-Dzyaloshinskii-Moriya-interaction (DMI) model. Due to their
relativistic origin, these parameters are 
responsible for the electric polarization induced in the presence of a 
non-collinear spin alignment in materials with a centrosymmetric
crystal structure. Beyond this, our approach gives direct access to
the element- or 
site-resolved electric polarization. To demonstrate the capability of
the approach, we consider several examples of the so-called type II
multiferroics, for which the magneto-electric effect is observed either
as a consequence of an applied magnetic field (we use Cr$_2$O$_3$ as a
prototype), or as a result of a phase transition to a spin-spiral
magnetic state, as for instance in MnI$_2$, CuCrO$_2$ and AgCrO$_2$.
\end{abstract}

\maketitle


\section {Introduction}

The growing interest in multiferroic materials is motivated by the
possibility to control their magnetic or electric polarization by
applying external electric or magnetic fields, respectively
\cite{SC82,Kho09,PZ12,OKSK15}. This allows various technical
applications in electronics, spintronics, etc.\ \cite{LCS21}.
In particular, applying a static magnetic or electric field to control
the properties of multiferroic materials, one can minimize the energy
losses caused by induced electric currents.
This feature is a consequence of coexisting ferroelectricity (FE) and
ferromagnetism (FM) in multiferroic systems, which obey 
individually different symmetry requirements, either broken space
inversion symmetry ${\cal I}$ or broken time reversal symmetry ${\cal
  T}$, respectively, however, with the combined ${\cal I T}$ symmetry
being conserved.

While there are different types of multiferroics discussed in the
literature (see, e.g. \cite{CM07,Kho09,PZ12,TSN14,BC23}),
we will focus here on the materials which belong to the so-called type-II
multiferroics, in which electric polarization appears only in the
magnetically ordered phase due to a strong coupling between magnetism
and ferroelectricity. Any perturbation in these materials, breaking 
time reversal symmetry ${\cal T}$, gives rise to an electric
polarization being allowed
as a result of broken inversion symmetry ${\cal
  I}$ \cite{Mos06,Kho09,MNN11,Sol25}. This occurs in collinear
antiferromagnetic (AFM) systems, where an electric polarization can be
induced by an applied magnetic field, as it is observed, for instance, in 
$\beta$ Cr$_2$O$_3$ \cite{Ast61,MCSV12,BLQ+24}. In this case,
the linear magneto-electric (ME) effect is described by the ME tensor
$\underline{\alpha}$ giving the induced electric polarization $\vec{P}$
via the relationship $\vec{P} = \underline{\alpha}\vec{H}$.
As an example for materials exhibiting the spontaneous ME effect, one can mention the
triangular-lattice AFM compounds (TLA) such as MnI$_2$, CuFeO$_2$,
CuCrO$_2$, AgCrO$_2$, etc. These materials undergo a phase transition at low 
temperature to a non-collinear magnetic state
\cite{YOK+10me,FEP+12,POL+16,OMY+94,LOM+11}, as a consequence of the
competition of the frustrated AFM exchange interactions. 
In this work, however, we do not discuss the mechanisms responsible for
the stabilization of the AFM state, focusing instead on the electric
polarization for different non-collinear magnetic structures.

Despite common symmetry requirements, the microscopic mechanism
for the ME effect in the type II multiferroics may differ 
for different materials \cite{MNN11,Kur20}. 
The spin-orbit driven mechanism suggested by Katsura, Nagaosa and
Balatsky (KNB) was interpreted in terms of the spin current
\cite{KNB05} appearing between the two spin moments, $\hat{s}_1$ and
$\hat{s}_2$, due to their non-collinear alignment. This theory 
 suggests the rule that the induced electric
 polarization is characterized by the expression $\vec{P}_{12} \sim  
{\vec e}_{12} \times (\hat{s}_1 \times \hat{s}_2)$ with ${\vec
  e}_{12}={\vec   R}_{12}/|{\vec R}_{12}|$ (${\vec R}_{12}$ is the 
vector connecting the two sites). This rule indeed explains the trend of
the ferroelectric behavior for a large number of multiferroics with a
cycloidal-type spiral spin order \cite{KGS+03,XWWS08,TAT+06}. A similar
rule was suggested by Mostovoy \cite{Mos06}, that was obtained by applying the
phenomenological theory of inhomogeneous ferroelectric magnets to systems 
with cubic symmetry.  
  However, this rule fails to explain the spin-driven ferroelectricity
  for instance in triangular-lattice multiferroic helimagnets
  \cite{SOT08,Ari07,WCX+12}, as a consequence of the special symmetry used
  in these works, implying a cubic environment for the magnetic dimers
  (see, e.g. Refs.\ \cite{TS10,TSN14,Sol25}).
  For that reason, other mechanisms were 
suggested to explain the observed magneto-electric effect in these
materials. For instance, the electric 
polarization in CuFeO$_2$ having a helimagnetic structure was 
attributed to the single-ion effect due to the spin-dependent $p-d$
hybridization \cite{Ari07,TSN14}.

To go beyond the phenomenological model treatment of the ME effect, a
more general approach based on density functional theory (DFT)
calculations has been used 
\cite{XWWS08,XKZ+11,Sol17}. However, a direct DFT-based treatment of the
extended spin structures (e.g. long-period spin spirals) requires
supercell calculations, which are in general rather time-consuming. A
simplification can be achieved by the 
parametrization scheme given for instance by Xiang et al. \cite{XKZ+11}
within the generalized KNB model representing the spin-induced polarization
  in terms of spin-dimer contributions. This approach could be applied
  successfully to describe the induced electric polarization 
  of various systems with complex spin textures  \cite{YLL+12,CCCK18,EBPS25}.
  Another generalized KNB formalism has been reported by Solovyev
  \cite{Sol17}, which is based on a parametrization using the Berry-phase theory for the effective
  half-filled Hubbard model. It was successfully 
applied to describe the spin-induced electric polarization in different
multiferroic materials \cite{Sol17,NS19,SON21,Sol25}. 
Note that these approaches represent the 'pure
electronic' contribution to the induced polarization.
 An alternative mechanism was considered by Sergienko and Dagotto
   \cite{SD06}, to explain the electric polarization in perovskite
   RMnO$_3$ (R= Gd, Tb, Dy) having modulated magnetic structure.
It postulates an electric polarization appearing due 
to the spin-orbit-coupling (SOC) driven atomic displacements induced by
the non-collinear spin alignment.  
This is the so called 'lattice mediated polarization' mechanism
which is discussed in terms of the Dzyaloshinskii-Moriya  
interaction and may be responsible for the stabilization of helical
magnetic order in the presence of atomic displacements that break the
inversion symmetry in the system. Accordingly, it is referred to in the
literature as the inverse-DMI mechanism.  
Xiang et al. \cite{XWWG13} extended their generalized KNB theory by
taking into account the ion-displacement contribution to the
spin-order induced electric polarization.
Its importance for the multiferroics LiCuVO$_4$, TbMnO$_3$, and
HoMnO$_3$ (exhibiting spin spirals) was then discussed in relation to the electronic
contribution. Also in the case of the collinear AFM 
Cr$_2$O$_3$  the 'lattice mediated polarization' represents a dominating 
contribution to the magneto-electric polarization, as it was
demonstrated by different groups \cite{Ini08,YV14,BLQ+24}. 

We  will focus in the following on the 'pure electronic' effect, using the approach
described in the  Section \ref{Section:theory}, representing the
parametrization of the electric polarization in full analogy to the
parametrization of the spin-lattice Hamiltonian
\cite{MPL+22,MLPE23}. The corresponding parameters 
are calculated from first principles via the Green function (GF)
technique implemented using the Korringa-Kohn-Rostoker
(KKR) band structure method. In the following sections we apply this approach to
different materials demonstrating the properties which can be discussed
on the basis of such calculations. 
 It is worth noting that this approach can be
  extended accounting for the lattice-mediated polarization mechanism.
  This can be achieved by making use of our previous development of a scheme
  to calculate the spin-lattice coupling constants \cite{MPL+22,MLPE23}.
  This issue is briefly considered in Section \ref{Section:theory} and
  in the Section \ref{Section:summary}, but is
  beyond the aims of the present work.

\section{Theoretical details} \label{Section:theory}

In a recent study \cite{MPL+22,MLPE23} we discussed the spin-lattice
interaction represented by the coupling tensors 
$\underline{J}_{ij,k}$ which allows to get the force on an atom at $\vec{R}_k$ 
generated due to the perturbation caused by the rotation of the spin
moments $\hat{s}_i$ and $\hat{s}_j$ on sites $i$ and $j$, respectively.
Using the same idea for the magnetically induced electric polarization
$\Delta \vec{P}$, we start with the definition
\begin{eqnarray}
  \Delta \vec{P} &=& \frac{1}{\Omega}\int d^3r\, e \vec{r}\, \Delta \rho(\vec{r}) \; .
\label{eq:dipol_1}
\end{eqnarray}
In the case of a solid with $N_{u.c.}$ atoms per unit cell which are specified by the basis
vectors $t_q$ ($q = 1,N_{u.c.}$) and have an associated volume $\Omega_{q}$,
the integral in Eq.\ (\ref{eq:dipol_1}) can be written as follows
\begin{eqnarray}
  \Delta \vec{P} &=& \frac{e}{\Omega}\sum_n\sum_{q}\int_{\Omega_q} d^3r
  [\vec{r}_q+\vec{R}_n+\vec{t}_q] \Delta
  \rho(\vec{r}_q)\nonumber \\
      & = &\frac{e}{\Omega}\Bigg[\sum_n \sum_{q}\int_{\Omega_q} d^3r_q
  \vec{r}_q \Delta \rho(\vec{r}_q)  \nonumber \\
      &  & + \sum_{n} \vec{R}_n \sum_{q}  \int_{\Omega_q} d^3r_q
  \Delta \rho(\vec{r}_q) \nonumber \\
      &  & + \sum_{n} \sum_{q} \vec{t}_q
   \int_{\Omega_q} d^3r_q
  \Delta \rho(\vec{r}_q)\Bigg]  \; .
\label{eq:dipol_2}
\end{eqnarray}
Here $\vec{R}_{n}$ is a Bravais vector that describes the position
of the unit cell, and $\vec{r}_q = \vec{r}-\vec{R}_n-\vec{t}_q$ are atom
cell centered coordinates.
The  second term on the right-hand side of Eq.\ (\ref{eq:dipol_2})
vanishes due to charge neutrality of the unit cell.
The third term describes the change of the electric polarization of the unit cell
due to a charge redistribution within the cell \cite{GT66}. This term is not  
discussed here, as we focus on the polarization due to a
charge distortion under the assumption of charge
conservation for each atom.
 Furthermore, the atomic positions described by the Bravais vectors
  $\vec{R}_{n}$ are assumed to be fixed, i.e. any atomic displacements
  are neglected. This implies that the lattice mediated contributions to
  the induced electric polarization (see, e.g. \cite{SD06,Ini08, YV14}) are
  not taken into account.
  Accordingly, a direct comparison between the theoretical
  results presented below  to experimental data is not always meaningful.
  Due to the restriction made, the expression given above can be
  simplified to the form   
\begin{eqnarray}
  \Delta \vec{P}
  &\approx& \sum_{q}  \frac{1}{\Omega_{\rm u.c.}} \int_{\Omega_{q}} d^3r_q \vec{r}_q
  \Delta \rho(\vec{r}_q)  = \sum_{q} \Delta \vec{P}_q  \; ,
\label{eq:dipol_3}
\end{eqnarray}
giving the induced electric polarization of a unit cell after summation
over the induced electric polarization of all atoms in the cell. This is seen as an on-site 
approximation assuming a pronounced localization of the electronic
states contributing to the electric polarization of the atoms.
As a consequence, for each site $q$ we will discuss the  electric polarization 
evaluated with respect to the site origin and normalized by the unit
cell volume, which is induced by all contributions associated with spin
rotations on the surrounding atoms $i$ and $j$.

 Although the modern theory of electric polarization is more
  adequate for periodic 
  systems \cite{VK93,KV93,Res93}, Eq. (\ref{eq:dipol_1})
  seems appropriate to describe the polarization in an atomistic
  real-space formulation, 
similar to that used in other works based on the generalized KNB model
\cite{JONH06,XKZ+11,Sol17,EBPS25}. 
The present approach is based on the multiple scattering Green function 
formalism to represent the electronic structure, which allows
to avoid the use of the Bloch theorem for the wave functions. 
As a natural consequence, the induced charge density  $\Delta
\rho(\vec{r}_q)$ in Eq.\ (\ref{eq:dipol_3}), as well as the corresponding
induced polarization, will be formulated below as a real space
summation over non-periodic contributions induced around 
distorted spin dimers (i.e., the tilted spin moments). This implies, the
real-space formulation given by Eq.\ (\ref{eq:dipol_1}) may be justified
by using the correspondence to finite systems, e.g. clusters or
molecules. Obviously, the applicability of 
this approach is determined first of all by the approximations used to
go from the expression given by Eq. (\ref{eq:dipol_1}) to that given
by Eq.\ (\ref{eq:dipol_3}).

In general, Eq.\ (\ref{eq:dipol_3}) obviously may represent the change of the local
polarization on lattice site $q$ as a response to a perturbation
induced by a rotation of magnetic moments.
Here we restrict ourselves to systems with a centrosymmetric crystal
lattice and invariant with respect to time reversal in its reference state.
This implies that a finite electric polarization may occur only as a
consequence of spin rotations creating a non-collinear magnetic structure.

In terms of the electronic Green function the electric polarization can be written as follows 
\begin{eqnarray}
  \Delta \vec{P} &=& -\frac{1}{\pi} {\rm Im}{\rm Tr}\int dE
  \frac{1}{\Omega_{\rm u.c.}}  \nonumber \\
  &&\times \sum_q \int_{\Omega_q} d^3r_q (e\vec{r}_q) \Delta G(\vec{r}_q,\vec{r}_q,E) \; .
\label{eq:dipol_4}
\end{eqnarray}

In the presence of a perturbation $\Delta V (\vec{r})$,
the corresponding change of the Green function occurring in
Eq.\ (\ref{eq:dipol_4}) is given up to second order by
\begin{eqnarray}
  \Delta G(E) & = &  G_0(E) \Delta V G_0(E) \nonumber \\
  && + G_0(E) \Delta V G_0(E)
\Delta V G_0(E) + ...  \; ,
\label{eq:Delta_GF}
\end{eqnarray}
with $G_0(E)$ the Green function of the system in its unperturbed
reference state.
For the sake of simplicity, we omit the arguments  $\vec{r}$ for
the potential and Green functions, keeping in mind an integration
over the region of the applied perturbation. 
In the case of perturbations due to spin tiltings on two  lattice sites $i$ and
$j$ only the second term in Eq.\ (\ref{eq:Delta_GF}) gives a
  non-vanishing contribution to the electric polarization, leading to the expression
\begin{eqnarray}
  \Delta \vec{P} &=& -\frac{1}{\pi} \frac{e}{\Omega_{\rm u.c.}}{\rm Im\,
    Tr} \int dE \sum_q  \nonumber \\
 &&  \int_{\Omega_{q}}  d^3r_q  \int  d^3r'  \int  d^3r''
       \vec{r}_q \; G(\vec{r}_q,\vec{r}\,',E) \Delta V_i(\vec{r}\,')
       \nonumber \\
  &&   \times             G(\vec{r}\,',\vec{r}\,'',E) \Delta V_j(\vec{r}\,'')
                       G(\vec{r}\,'',\vec{r}_q,E) \; . 
\label{eq:dipol_5}
\end{eqnarray}
With this, one can introduce the parameters $p_{ij,k}^{\alpha \beta, \mu}$ defined as derivatives with 
respect to the spin direction on sites $i$ ($\vec{R}_i =\vec{R}_{n_i}
+\vec{t}_{q_i}$) and $j$ ($\vec{R}_j =\vec{R}_{n_j} +\vec{t}_{q_j}$),
$\Delta s^\alpha_i$ and $\Delta s^\beta_j$, respectively, giving the
polarization $\vec{P} = (P^x,P^y,P^z)$ on site $k$ with $\vec{R}_k = \vec{t}_{q_k}$ in
the unit cell $\vec{R}_{n_k} = 0$, as follows  
\begin{eqnarray}
  \Delta {P}^\mu &=&  \sum_{i,j,k} \frac{\partial^2 (\Delta
    {P}_k^{\mu})(\{\hat{s}_n\})}{\partial s^\alpha_i \partial s^\beta_j} \Delta
  s^\alpha_i \Delta s^\alpha_j  \nonumber \\
    &=& \sum_{i,j,k} {p}_{ij,k}^{\alpha \beta, \mu} \Delta
  s^\alpha_i \Delta s^\alpha_j \,, 
\label{eq:dipol_6}
\end{eqnarray}
 with $\mu$ indicating the Cartesian coordinates $x, y$ and $z$.
Later on we will distinguish symmetric and antisymmetric terms with respect
to a permutation of the Cartesian indices $\alpha$ and $\beta$.

Similar to the treatment of the spin-lattice coupling parameters \cite{MPL+22,MLPE23}
the KKR-GF formalism allows to express the parameters $\vec{p}\,_{ij,k}^{\alpha \beta} = ({p}_{ij,k}^{\alpha \beta,x},{p}_{ij,k}^{\alpha \beta,y}, {p}_{ij,k}^{\alpha \beta,z})$
 in terms of the KKR scattering path operator matrix $\underline{\tau}_{i
  j}(E)$ as follows
\begin{eqnarray}
  \vec{p}\,_{ij,k}^{\alpha\beta} &=& - \frac{1}{2\pi} \frac{1}{\Omega_{\rm u.c.}} 
 {\rm Im\, Tr} \int dE \nonumber \\
&&\times  \bigg[ \langle Z^k | e\vec{r} | Z^k \rangle  \underline{\tau}_{ki}
    \underline{T}_i^\alpha \underline{\tau}_{ij}  \underline{T}_j^\beta
    \underline{\tau}_{jk} \nonumber \\
&&+  \langle Z^k | e\vec{r} | Z^k \rangle \underline{\tau}_{kj} \underline{T}_j^\beta
    \underline{\tau}_{ji} \underline{T}_i^\alpha  \underline{\tau}_{ik}  \bigg] \; ,
\label{eq:dipol_KKR}
\end{eqnarray}
with $\hat{\tau}_{ij}(E)$ standing for the 
scattering path operator that transfers an electronic wave coming in at 
site $j$ into a wave going out from site $i$ with all possible
intermediate scattering events accounted for.

One has to stress that the parameters $\vec{p}\,_{ij,k}^{\alpha\beta}$ are defined using the same normalization 
as the induced electric polarization of the unit cell, i.e. both quantities
are normalized w.r.t. the volume of the unit cell (see Eqs.\ (\ref{eq:dipol_5}) and
(\ref{eq:dipol_KKR})). With this definition the electric polarization
of the unit cell is obtained directly by a summation over the atomic
contributions (note that by definition these have to be distinguished from 
the atomic dipole moments).   
Using the relativistic spin-angular representation \cite{Ros61},
the elements of the matrix  $\underline{T}_i^\alpha$ are given by the expression 
\begin{eqnarray}
T^{\alpha}_{i,\Lambda'\Lambda}
& = & \int_{\Omega_{q}} d^3r\; Z^{i\times}_{\Lambda'}(\vec{r},E)\;\left[\beta
\sigma_{\alpha}B_{\rm xc}({r})\right] Z^{i}_{\Lambda}(\vec{r},E)\, ,\nonumber \\
\label{matrix-element1}
\end{eqnarray}
and for the elements of the matrix $\langle Z^k | e\vec{r} | Z^k \rangle$ one has
\begin{eqnarray}
 \langle Z^k | e\vec{r} | Z^k \rangle_{\Lambda'\Lambda}
& = & \int_{\Omega_{q}} d^3r\;
Z^{k\times}_{\Lambda'}(\vec{r},E)\left(e \vec{r}\right) Z^{k}_{\Lambda}(\vec{r},E)\, ,\nonumber \\
\label{matrix-element2}
\end{eqnarray}
where the four-component wave functions
$Z^{i}_{\Lambda}(\vec{r},E)$ stand for the regular
solutions of the single-site Dirac equation with the Hamiltonian set up
within the framework of relativistic spin-density functional theory
\cite{MV79,ED11,EBKM16} with $\Lambda = (\kappa,\mu)$, and
$\kappa$ and $\mu$ being the spin-orbit and magnetic quantum numbers,
$\beta$ is one of the  standard Dirac matrices 
\cite{Ros61} and $B_{\rm xc}({r})$  
is the spin dependent part of the exchange-correlation potential. Note that the
energy dependence of the scattering 
path operator and of the matrices $\underline{T}_{i}^\alpha$  is omitted in Eq.\ (\ref{eq:dipol_KKR}).

By making a decomposition of the polarization parameters into symmetric
${p}^{\nu,{\rm s}}_{ij,k}$ and antisymmetric $\vec{p}\,^{\nu,{\rm a}}_{ij,k}$
parts, one gets access to the following parametrization of the electric
polarization as suggested by Moriya \cite{Mor68} and used later on in
generalized KNB theories \cite{XKZ+11,Sol17,Sol25}, which is induced by a
spatial spin modulation,  
\begin{eqnarray}
  \Delta{P}^\nu_k &=& \sum_{ij} \Bigg[  {p}^{\nu,{\rm s}}_{ij,k} \{(\hat{s}_i
    \cdot \hat{s}_j) - \hat{s}^z_i\hat{s}^z_j)\} \nonumber \\
   && +   \vec{p}\,^{\nu,{\rm a}}_{ij,k} \cdot (\hat{s}_i \times \hat{s}_j) +
\hat{s}_i {\underline \Pi}^{\nu}_{ii,k} \hat{s}_i\Bigg] \,
\label{Eq:parametrized_polarization}
\end{eqnarray}
with $\vec{p}\,^{\nu,{\rm a}}_{ij,k} = ({p}^{x \nu,{\rm
    a}}_{ij,k}, {p}^{y \nu,{\rm a}}_{ij,k}, {p}^{z \nu,{\rm
      a}}_{ij,k})$ and ${p}^{\gamma \nu,{\rm a}}_{ij,k} =
  \frac{1}{2}\epsilon_{\alpha\beta\gamma} p\,_{ij,k}^{\alpha \beta, \nu}$, where
  $\epsilon_{\alpha\beta\gamma}$ is the Levi-Civita symbol and $\gamma = {x,y,z}$.       
This approach allows now a general description of the ME effect, giving
in particular access to the properties of various compounds avoiding the restrictions caused
by phenomenological Hamiltonians (see also discussions by
Solovyev et al. \cite{Sol25}).
This concerns, in particular, the model used to describe the
ME effect depending exclusively on the leading exchange mechanism (i.e.\ direct
exchange, superexchange, etc.), while for some materials the 
exchange interactions may be determined by several mechanisms competing
with each other. For instance, two competing types of Cr-Cr
exchange interactions are discussed in the case of the TLA compounds
$Me$CrO$_2$, namely, the FM superexchange and AFM direct
exchange \cite{AD91,Maz07,UKYK13}. 

Furthermore, it is pointed out, that the parameters ${p}^{\nu,{\rm
    s}}_{ij,k}$ and $\vec{p}\,^{\nu,{\rm a}}_{ij,k}$ 
are, in general, three-site parameters. They supply the information
concerning the individual electric polarization of different atoms $k$
beyond the special case of $k=i$ and $k=j$ with $i$ and $j$
being the site indices for the 
magnetic atoms with their spin moment rotated.

 In the next sections we will focus on the SOC-driven polarization
 described by the parameters antisymmetric with respect to spin permutation,
$\vec{p}\,^{\nu,{\rm a}}_{ij,k}$. These parameters
 are closely connected to the DMI-like spin-lattice coupling (SLC)
 parameters discussed in Ref.\ \onlinecite{MPL+22,MLPE23}, which are 
responsible for the forces ${\cal
  F}^\mu_k = -\sum_{\alpha,i,j} {\cal  D}^{\alpha,\mu}_{ij,k} (\hat{e}_i \times 
\hat{e}_j)_\alpha$ on an atom $k$ in the lattice, arising due to spin
canting on sites $i$ and $j$. These 
forces appear as a consequence of the electric polarization induced on
the sites $k$ (see Eq.\ (\ref{eq:dipol_2})). 
Their competition with the elastic interatomic forces may lead
(as discussed in Ref.\ \onlinecite{SD06,XWWG13}), or may not lead to atomic
displacements. In the former case, the induced local polarization
  occurs due to the ion displacements and can be treated as a
  lattice-mediated contribution to the generalized inverse-DMI
mechanism. This mechanism is not taken into account in the 
present work, and accordingly we discuss only the spin-induced polarization appearing
in the case of undistorted atomic positions.

\section {Computational details}

All calculations are performed using the fully relativistic KKR
Green function method \cite{SPR-KKR8.5,EKM11}.
The exchange-correlation potential is determined within the 
framework of the general gradient approximation (GGA) to density
functional theory (DFT), using the Perdew-Burke-Ernzerhof (PBE)
parametrization \cite{PBE96}. 
A cutoff $l_{max} = 3$ was used for the angular momentum expansion of 
the Green function. Further details can be found for example in
Refs.\ \onlinecite{MLPE23} and \onlinecite{MPE20}.
Integration over the Brillouin zone (BZ) has been performed using a
$28 \times 28 \times 15$ $\vec{k}$-mesh for  MnI$_2$, a  $22 \times
22 \times 22$ $\vec{k}$-mesh for CuCrO$_2$ and a $20 \times 20 \times
20$ $\vec{k}$-mesh for Cr$_2$O$_3$. 

The electronic structure and electric polarization parameters
for the  MnI$_2$ and CuCrO$_2$ compounds have been calculated within the
full-potential approach, i.e.\ without restrictions to the shape of the electron
potential.
This scheme implies a space separation by planes dividing the bonds between
neighboring atoms, drawn perpendicular to the interatomic
vectors. The planes are positioned to divide the bonds in proportion to the
atom-dependent radii evaluated using the cross points of Hartree
potentials of neighboring atoms. This modified Wigner-Seitz construction
gives a space separation very similar to a Bader construction that,
however, is much more elaborate. 
For technical reason, the calculations for Cr$_2$O$_3$ have
been performed using the atomic sphere approximation (ASA) with the
Wigner-Seitz radii $R_{WS}^{Cr} = 2.7$ a.u. and $R_{WS}^{O} = 2.3$
a.u. for Cr and O atoms, respectively.

\section {RESULTS}

In the following we present few representative examples for calculations on the ME effect
to demonstrate the application of our approach to several prototype
multiferroic systems, comparing the results with the available
experimental and theoretical data. The calculations were performed for the
triangular-lattice antiferromagnets MnI$_2$ and CuCrO$_2$ with the spontaneous
ferroelectric effect induced by a helimagnetic spin structure and the
Cr$_2$O$_3$ collinear antiferromagnet exhibiting the linear magneto-electric
effect induced by an external magnetic field. Focusing on the so-called inverse-DMI 
polarization mechanism described by the second term in
Eq.\ (\ref{Eq:parametrized_polarization}), we achieve 
reasonable agreement of our results with those reported in the literature, 
obtained using the generalized KNB theory
in spite of our restriction to the electronic contribution to the spin
induced electric polarization \cite{XKZ+11,Sol17,Sol25}.

\subsubsection {MnI$_2$}

First, we consider the triangular-lattice antiferromagnet MnI$_2$.
It crystallizes in the CdI$_2$ type structure with centrosymmetric space
group $P\bar{3}m1$ (No. 164). Each Mn atomic layers forms a 2D
triangular lattice with AFM inter-atomic interactions leading to a
sequence of magnetic phase transitions upon cooling, first
at $T_{N1}= 3.95$ K  (to the non-collinear magnetic state with a proper
screw spin modulation), then 
at 3.8 K  ($T_{N2}$) and  3.45 K ($T_{N3}$ ) \cite{SKI95}. The in-plane
electric polarization of about 84 $\mu$C/m$^2$
emerges in the proper screw magnetic ground state below $T_{N3}$,
and is characterized by the propagation vector $\vec{q}_{m3} = (0.181,
0,0.439)$ \cite{KSI+11,WCX+12}.  
\begin{figure}[tb]
  \center
  \includegraphics[width=0.18\textwidth,angle=0,clip]{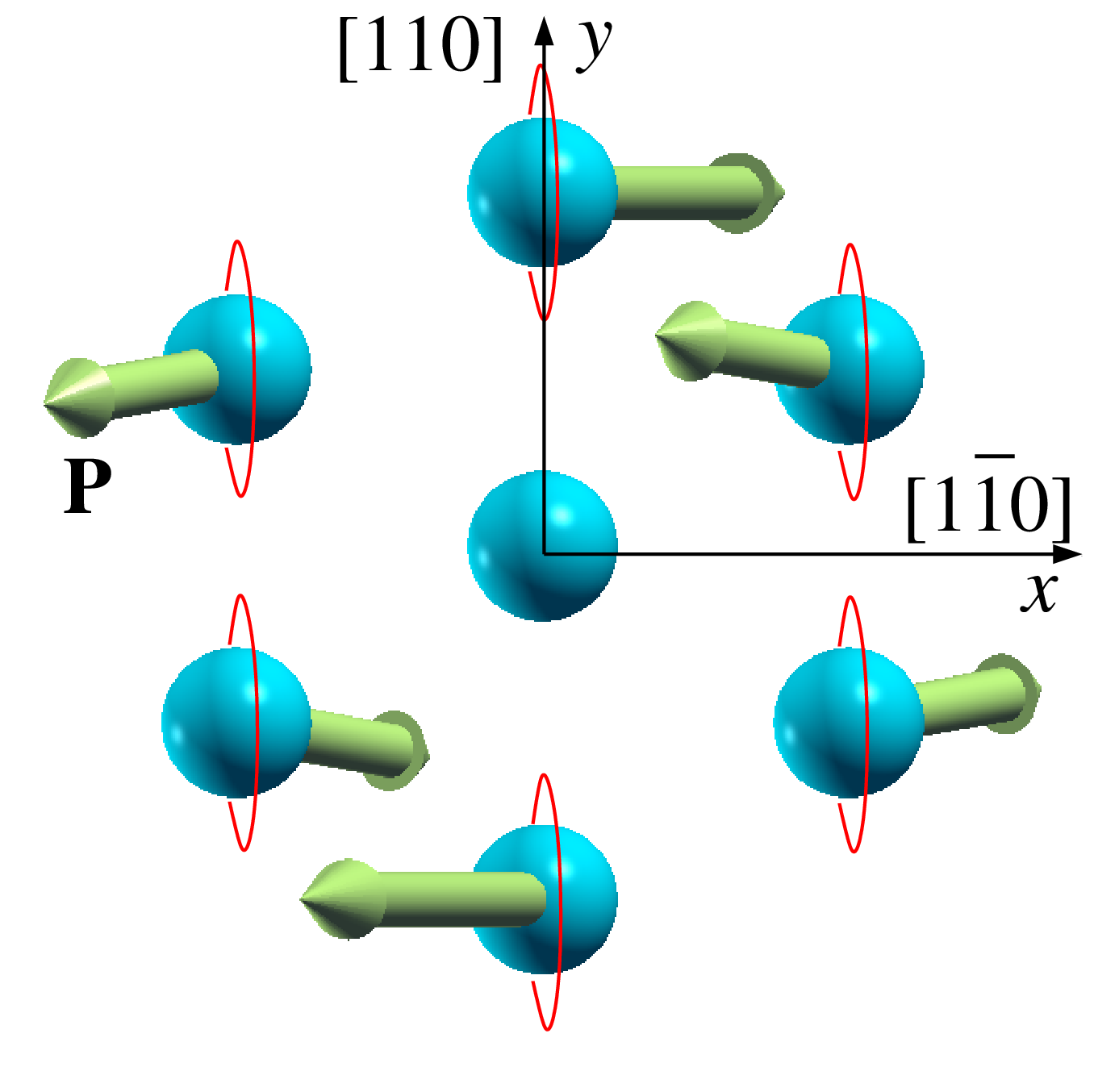}\,(a)
  \includegraphics[width=0.18\textwidth,angle=0,clip]{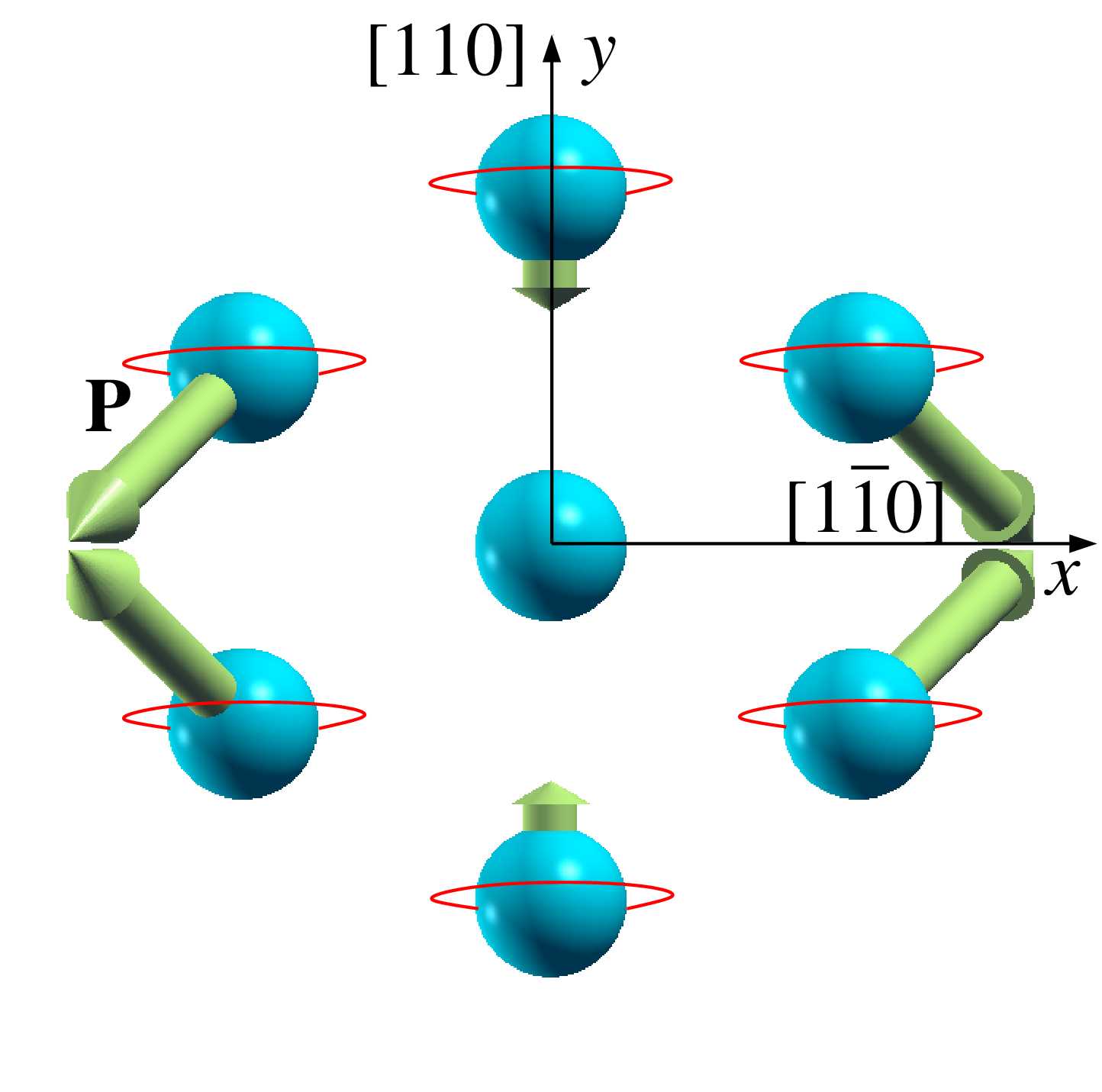}\,(b)
  \caption{\label{fig:I-DMI_MnI2-NN} The polarization vectors (shown by
    arrows) $\vec{p}\,^x_j = ({p}^{x x,{\rm 
    a}}_{0j,j},{p}^{x y,{\rm a}}_{0j,j},{p}^{x z,{\rm
    a}}_{0j,j}) $ (a) and  $\vec{p}\,^y_j = ({p}^{y x,{\rm
    a}}_{0j,j},{p}^{y y,{\rm a}}_{0j,j},{p}^{y z,{\rm
        a}}_{0j,j})$ (b) calculated for MnI$_2$, characterizing the
    contributions to the electric polarization due to spin rotation of
    the hearest-neighbor Mn atoms arranged around Mn$_0$. 
    Spin rotations occur within the  $yz$ (a) and $xz$ (b) planes shown
    by red circles.  
  }
\end{figure}

The parameters ${p}^{\gamma\nu,{\rm a}}_{ij,k}$ for MnI$_2$,
were calculated using Eq.\ (\ref{eq:dipol_KKR}) and working in the full-potential mode,
i.e. without restrictions for the shape of the electron potential and charge
density \cite{EBKM16}. 
Fig.\ \ref{fig:I-DMI_MnI2-NN} shows the vectors (a) $\vec{p}\,^x_j = ({p}^{x x,{\rm
    a}}_{0j,j},{ p}^{x y,{\rm a}}_{0j,j},{ p}^{x z,{\rm
    a}}_{0j,j}) $  and  (b) $\vec{p}\,^y_j = ({p}^{y x,{\rm
    a}}_{0j,j},{p}^{y y,{\rm a}}_{0j,j},{p}^{y z,{\rm
    a}}_{0j,j})$, respectively, characterizing the electric polarization $\vec{P}_j =
\vec{p}\,^x_j \cdot (\hat{s}_0 \times \hat{s}_j)_{x}$ and $\vec{P}_j =
\vec{p}\,^y_j \cdot (\hat{s}_0 \times \hat{s}_j)_{y}$ appearing due to the
relative rotation of the nearest-neighbor Mn spins for the atoms
Mn$_j$ arranged around the central atom Mn$_0$. 
The spin moments are rotated within the planes shown by red circles
and are assumed to be the same for every pair of atoms (Mn$_0$,
Mn$_j$). The spin rotations that break locally the inversion
  symmetry induce a site-dependent electric polarization for the Mn
  atoms, which depends on the crystal symmetry and the spin
  rotations, either within the $yz$ and $xz$ planes shown in figures (a) and (b), respectively
(a more detailed discussion of the symmetry properties can be found, for instance,
in Ref.\ \cite{Sol25}). 
In the case of periodic spin modulation, some polarization contributions
cancel each other. The total result depends on the type of spin spiral
as well as on the direction of propagation vector $\vec{q}$.

\begin{figure}[b]
\includegraphics[width=0.2\textwidth,angle=0,clip]{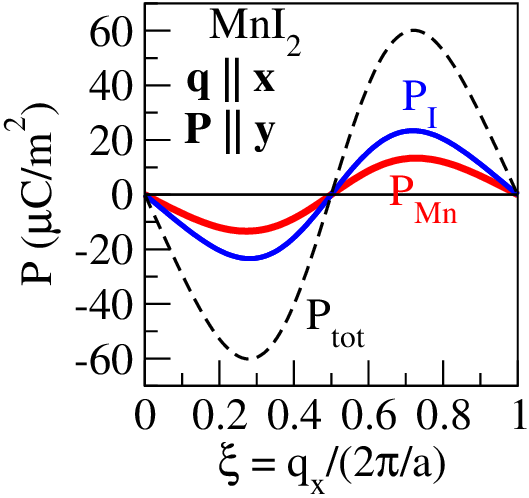}\,(a)
\includegraphics[width=0.2\textwidth,angle=0,clip]{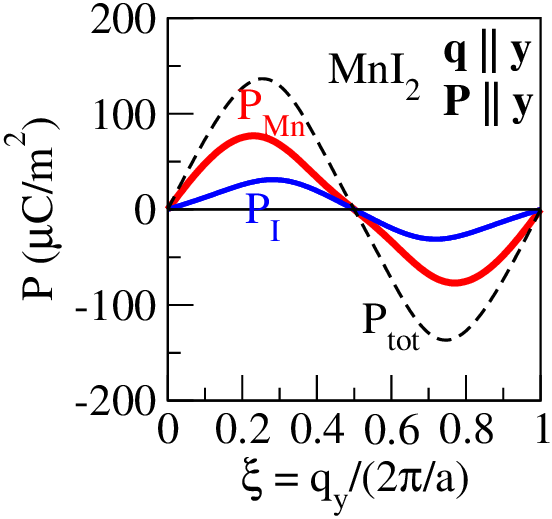}\,(b)\\
\hspace{4mm}
\includegraphics[width=0.17\textwidth,angle=0,clip]{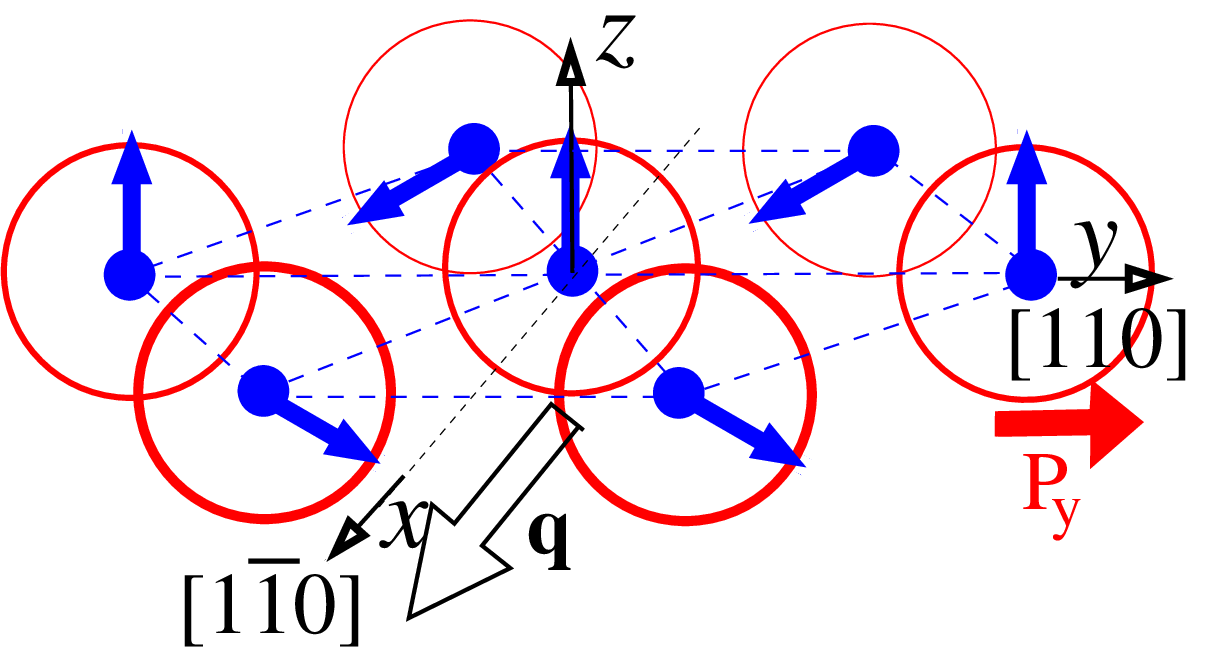}\,(c)
\hspace{5mm}
\includegraphics[width=0.17\textwidth,angle=0,clip]{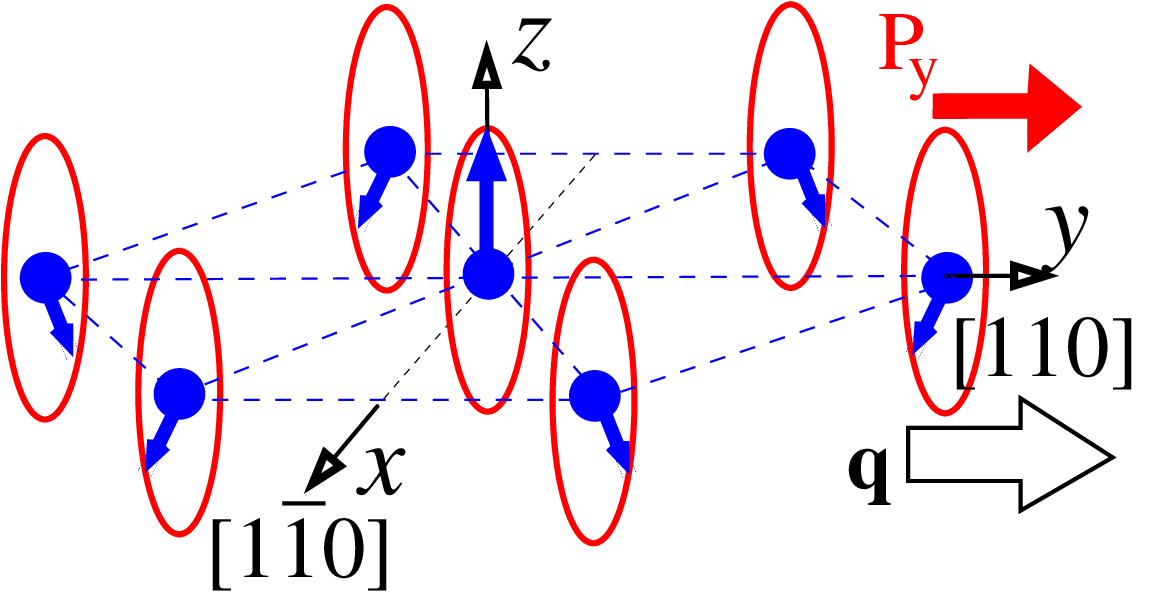}\,(d)\\
  \includegraphics[width=0.2\textwidth,angle=0,clip]{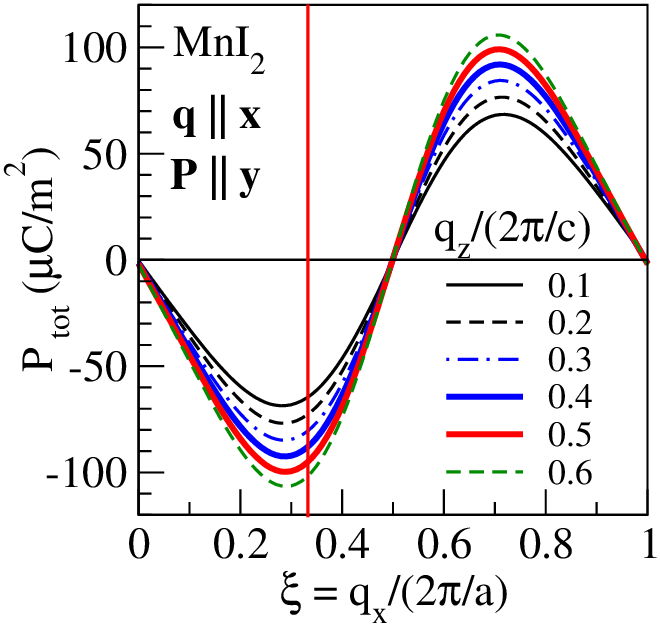}\,(e)
  \includegraphics[width=0.2\textwidth,angle=0,clip]{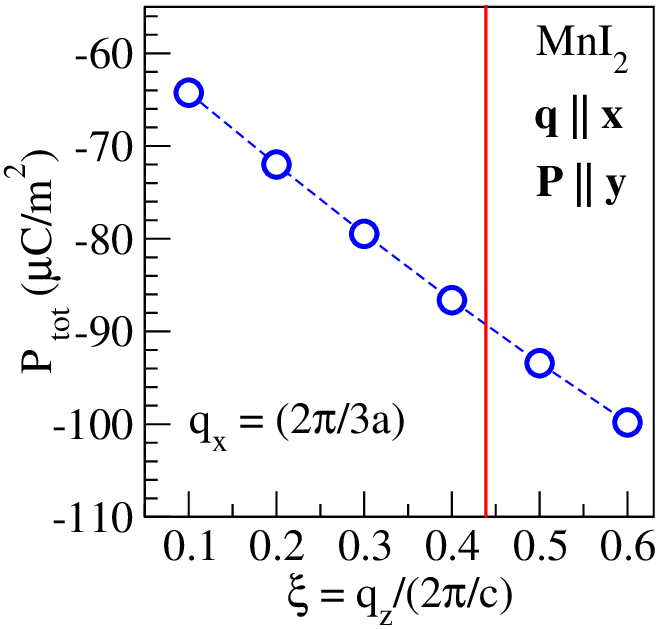}\,(e)
\caption{\label{fig:I-DMI_MnI2-qy} Element-resolved electric
  polarization $\vec{P}$ ($\mu$C/m$^2$) in MnI$_2$ due to proper
  screw magnetic structures characterized by propagation vectors
  $\vec{q} || \langle 1 \bar{1} 0 \rangle$ (a) and
  $\vec{q} || \langle 1 1 0 \rangle$ (b), represented as a function of
  the the propagation vector $\vec{q} = (q_x,0,0)$ and $\vec{q} =
  (0,q_y,0)$, respectively; (c) and (d) show shematically the spin
  modulation and the directions of the electric polarization $\vec{P}$,
  corresponding to (a) and (b), respectively; (e) Total electric
  polarization in MnI$_2$ due to proper screw magnetic structures with
  the out-of-plane propagation vectors  $\vec{q} = (q_x,0,q_z)$, with
  different curves corresponding to different values of $q_z$; (f) The
  dependence of the electric polarization on $q_z$ for the fixed value
  of the in-plane component of the propagation vector, $q_x = (2\pi/3a)$. }       
\end{figure}

Fig.\ \ref{fig:I-DMI_MnI2-qy} shows the polarization for the
proper screw spin modulation with the propagation vectors $\vec{q}_1 ||
[1 \bar{1} 0]$ (the $x$ direction) (a) and $\vec{q}_2 || [1 1 0]$ (the
$y$ directions) (b), plotted as a function of the in-plane 
propagation vectors $\vec{q}_1 = (q_x,0,0)$ and $\vec{q}_2 =
  (0,q_y,0)$, respectively.
It is non-vanishing in line with the predictions by the generalized KNB
models \cite{KSI+11,WCX+12,Sol17}.
The directions of the MnI$_2$ crystal lattice are characterized by
different symmetry: the $[1 1 0]$ crystallographic axis is a two-fold rotation axis
while the plane including the $[1 \bar{1} 0]$ and the $[0 0 1]$ axes (perpendicular to
the two-fold rotation axis) is a mirror plane.
 For both propagation vectors $\vec{q}_1$ and $\vec{q}_2$, the
 non-vanishing electric polarization is aligned along the $\langle 1 1 0 \rangle$
 crystallographic direction shown
schematically in Fig.\ \ref{fig:I-DMI_MnI2-qy} (c) and (d),
respectively. This is a consequence of symmetry  properties of the
system in the presence of spatial spin modulation,  which is discussed
in details in the literature (see, e.g., Refs.\ \cite{KSI+11,WCX+12,Sol17,Sol25}).

 As one can see in Fig.\ \ref{fig:I-DMI_MnI2-qy}, the electric
 polarization of  the Mn (blue line) and I (red line) sublattices are
 comparable in  magnitude and changes sign together with the orientation
of the propagation vector. Similar trends occur also for the total
 polarization shown in Fig.\ \ref{fig:I-DMI_MnI2-qy} by a dashed
 line; it has a maximum around ${q}_x = (2\pi/3a)$ (a) and ${q}_y
 = (2\pi/3a)$ (b).

 Although the propagation vector of the experimentally observed spin spiral
 has an out-of-plane component ${q}_{z} = 0.439(2\pi/c)$, we studied
 also the electric polarization for the proper screw spin spiral with
 different ${q}_{z}$ components of the propagation vector. The results are 
 shown in Fig.\ \ref{fig:I-DMI_MnI2-qy} (e) as a function of the
 in-plane propagation vector ${q}_{||} = (\xi, 0)(2\pi/a)$, where each
 curve corresponds to a different value of ${q}_{z} = \xi(2\pi/c)$.
 The red vertical line indicates the wave vector  ${q}_{x} = \xi(2\pi/3a)$.
 Fig.\ \ref{fig:I-DMI_MnI2-qy} (f) shows the dependence of the electric
 polarization on the parameter $\xi$ (characteerizing the component ${q}_{z} =
 \xi(2\pi/c)$) for the in-plane component  ${q}_x = 
 (1/3)(2\pi/a)$ (indicated by a red line in Fig.\ \ref{fig:I-DMI_MnI2-qy} (e)). In
 the case of ${q}_{z} = 0.44(2\pi/c)$ the induced electric 
 polarization is equal approximately to $90$ $\mu$C/m$^2$, which is rather close
 to the experimental value measured at $T = 2$~K as well as to 
 theoretical results \cite{WCX+12} accounting only for the electronic
 contribution to the polarization. In order to make a one-to-one
 comparison of their theoretical results (calculated for $T=0$K)  Wu et
 al. \cite{WCX+12} extrapolated the experimentally observed polarization to
 $T = 0$~K, leading to an electric polarization of about $128$
 $\mu$C/m$^2$. The deviation from the experimental value was attributed
 to the contribution due to the
 lattice mediated polarization, as was confirmed by corresponding calculations.

\subsubsection {CuCrO$_2$}

As a next example, we discuss the TLA compound CuCrO$_2$
seen as representative of the $Me$FeO$_2$ and 
$Me$CrO$_2$ oxides. For these materials a first-principles treatment of
a spin-induced electric polarization (i.e.\ beyond the
phenomenological model consideration) may be challenging because of the
competition of two mechanisms of nearest-neighbor exchange
interactions, i.e.\ the FM superexchange and AFM direct exchange.
One can expect a corresponding competition also for the magnetoelectric
polarization.  
At room temperature these materials crystallize in the delafossite
crystal structure having $R\bar{3}m$ space group symmetry.
Their magnetic properties are essentially determined by frustrated
antiferromagnetic exchange interactions \cite{YRH+06,POL+16}.
However, we do not discuss the origin of the magnetic ordering in
CuCrO$_2$, but only stress that the non-collinear AFM structure at
low temperature is accompanied by an emerging electric polarization,
similar to the TLA MnI$_2$ compound.  

\begin{figure}[b]
\includegraphics[width=0.2\textwidth,angle=0,clip]{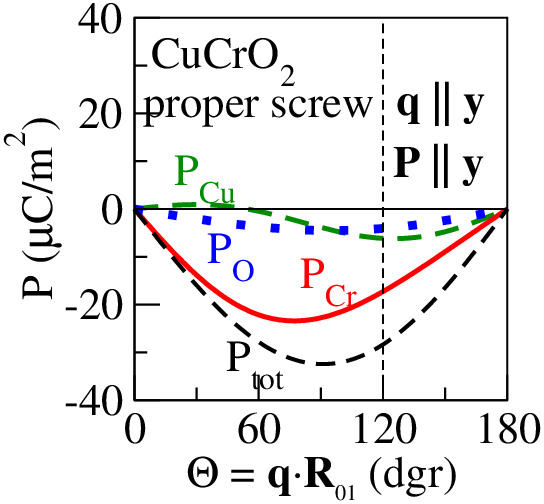}\,(a)
\includegraphics[width=0.2\textwidth,angle=0,clip]{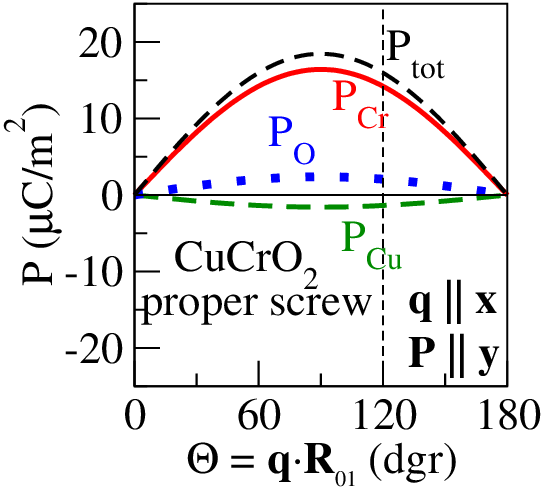}\,(b) \\
\hspace{4mm}
\includegraphics[width=0.17\textwidth,angle=0,clip]{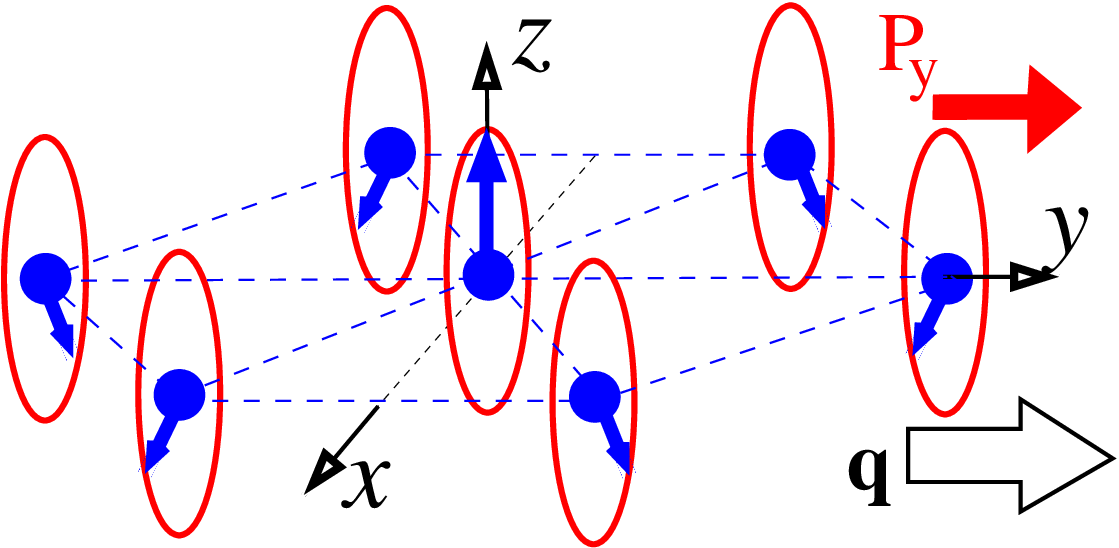}\,(c)
\hspace{5mm}
\includegraphics[width=0.17\textwidth,angle=0,clip]{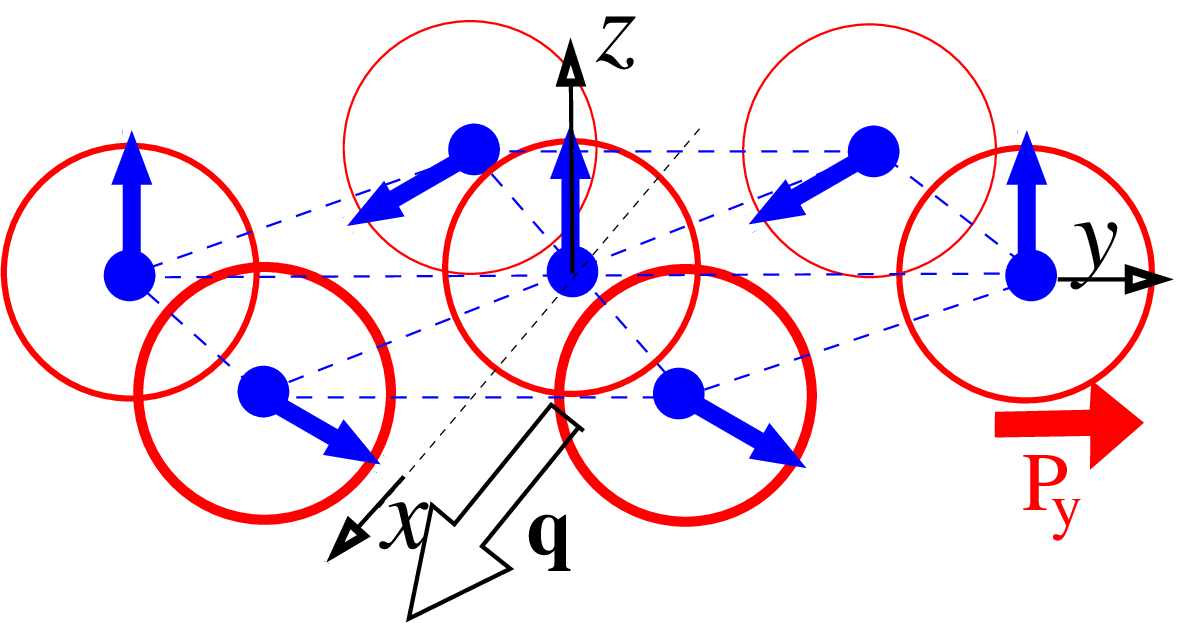}\,(d)
\caption{\label{fig:CuCrO2-polar_screw} The $\vec {p}^{\nu,{\rm a}}_{ij,k}$-related electric
  polarization for the CuCrO$_2$ compound: on Cu sites (green), Cr sites (red) and O sites (blue).
  Panels (a) and (b) show the polarization induced by a proper screw spin modulation
  characterized by the wave vectors $\vec{q}||\hat{y}||[110]$ and
  $\vec{q}||\hat{x}||[1\bar{1}0]$, respectively. In both cases a
  polarization emerges along the $y$ direction.
  The results are presented as a function of the angle  $\Theta =
  \vec{q}\cdot\vec{R}_{01}$ where $\vec{R}_{01}$ corresponds to the
  position of a nearest-neighbor Cr atom at $\vec{R}_{01} =
  (0.866,0.5,0)a$.
  The proper screw spin spiral in a 2D triangular Cr sublattice is shown
  schematically for $\vec{q}\,||\,\hat{y}\,||\,[110]$ (c) and for
  $\vec{q}\,||\,\hat{x}\,||\,[1\bar{1}0]$ (d), respectively.
  The red arrows show corresponding orientations of the induced electric 
  polarization.
}     
\end{figure}
Previous neutron diffraction studies on CuCrO$_2$ propose
 an out-of-plane 120$^o$ chiral spin structure below $T_N \sim 24$ K,
 as the ground state, with a propagation vector $\vec{q} = 
(1/3,1/3,0)$ \cite{KKA90} and incommensurate $\vec{q} = (0.329,329,0)$
 (either a proper screw or
 cycloidal) \cite{PDM+09},  while an incommensurate proper-screw magnetic
 structure was suggested in  Ref.\ \onlinecite{SKK+09}, that corresponds
 to a slightly modulated $120^o$ spin structure \cite{FEP+12}. 
The observed spontaneous electric polarization appearing below $T_N$ can be
attributed to the non-collinear magnetic structure indicating
the coupling between the ferroelectricity and spiral magnetic order
\cite{SOT08,KON+09,KNOK08,KNK+09,SKK+09}.
The proper screw spin modulation with the (110) spiral plane leads
  to the electric polarization $|\vec{P}| \approx 35$~$\mu$C/m$^2$ along the
  $[110]$ direction. An applied magnetic field (above 5.3 T) 
  $\vec{H}|| [1\bar{1}0]$ induces a transition to a cycloidal spin
  structure with a $(1\bar{1}0)$ spiral plane. This leads to a
  change of the orientation of the electric polarization towards $\vec{P} ||
  [1\bar{1}0]$ \cite{MFP+14} 
\begin{figure}[t]
\includegraphics[width=0.2\textwidth,angle=0,clip]{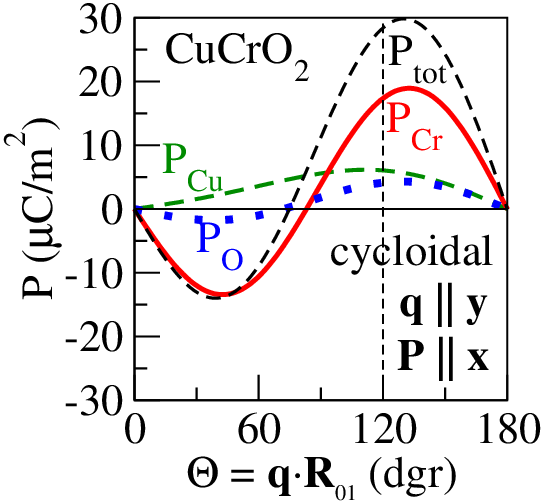}\,(a)
\includegraphics[width=0.2\textwidth,angle=0,clip]{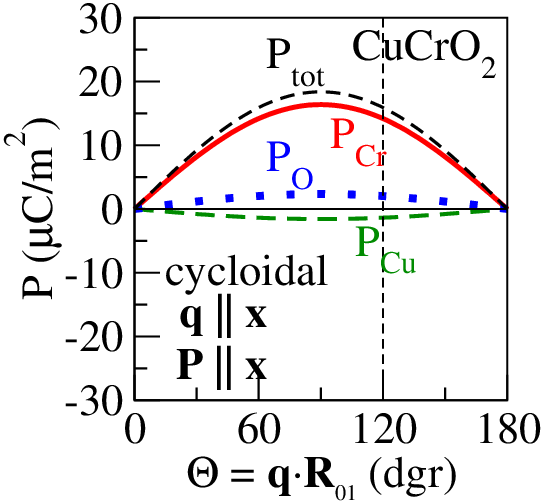}\,(b)\\
\includegraphics[width=0.2\textwidth,angle=0,clip]{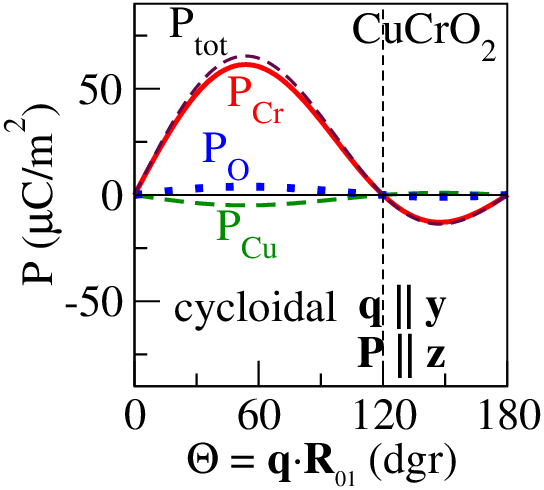}\,(c)
\includegraphics[width=0.2\textwidth,angle=0,clip]{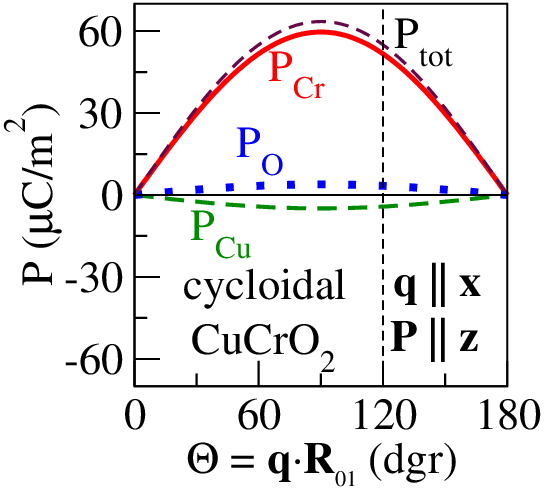}\,(d)\\
\hspace{4mm}
\includegraphics[width=0.17\textwidth,angle=0,clip]{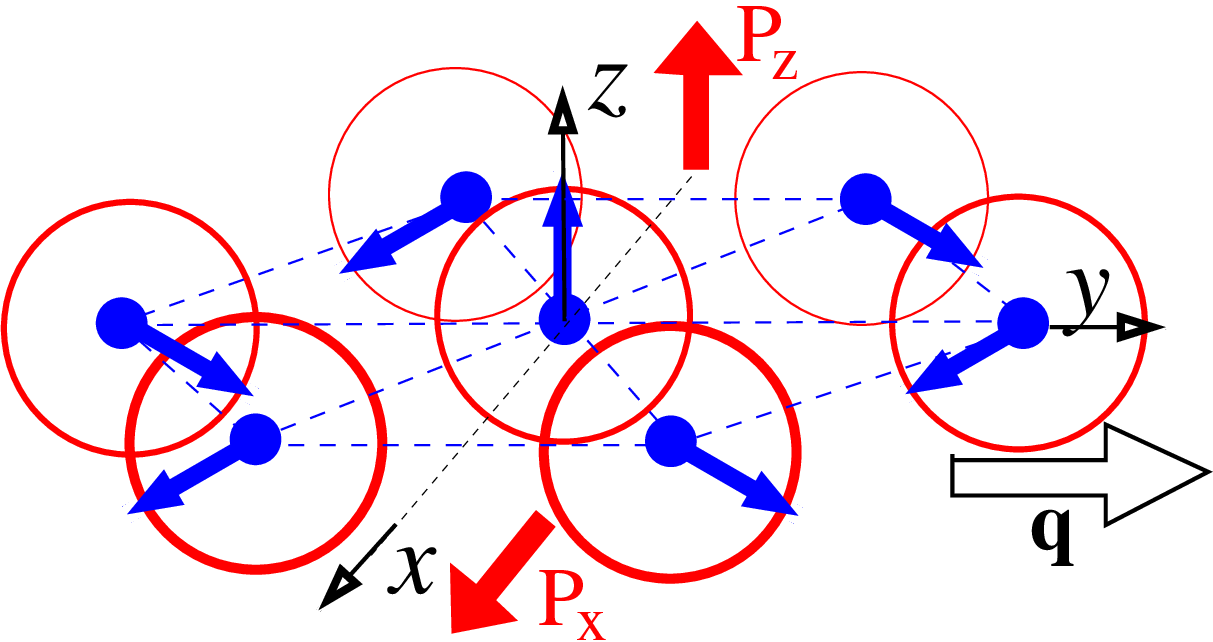}\,(e)
\hspace{5mm}
\includegraphics[width=0.17\textwidth,angle=0,clip]{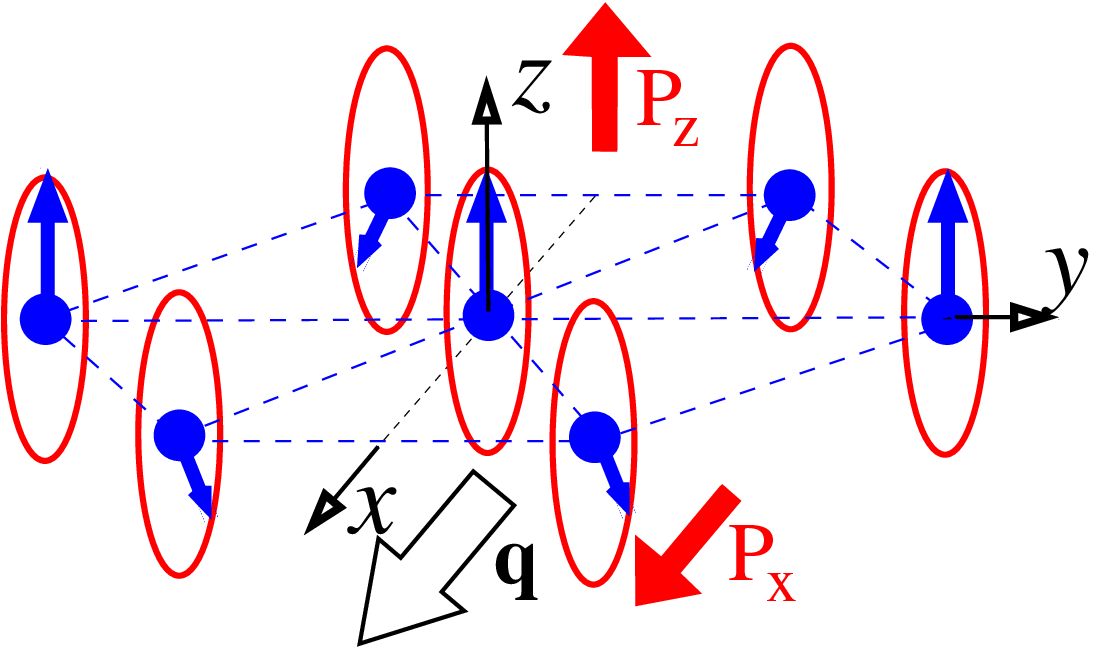}\,(f)\\
\caption{\label{fig:CuCrO2-polar_cycloid} The $\vec {p}^{\nu,{\rm a}}_{ij,k}$-related electric
  polarization for the CuCrO$_2$ compound: on Cu sites (green), Cr sites (red) and O sites (blue).
  Panels (a) and (c) show the $x$ and $z$ components of the
  polarization induced by a cycloidal spin modulation 
  characterized by the wave vector  $\vec{q}||\hat{y}||[110]$
  ($\vec{q}|| \hat{y}$). Panels (b) and (d) 
  show the $x$ and $z$ polarization components for a cycloidal spin
  modulation with $\vec{q}||\hat{x}||[1\bar{1}0]$ ($\vec{q}|| \hat{x}$),
  respectively. 
  The results are presented as a function of the angle  $\Theta =
  \vec{q}\cdot\vec{R}_{01}$ (similar to Fig. \ref{fig:CuCrO2-polar_screw}). 
  The cycloidal spin modulation in a 2D triangular Cr sublattice is shown
  schematically for $\vec{q}\,||\,\hat{y}\,||\,[110]$ (e) and for
  $\vec{q}\,||\,\hat{x}\,||\,[1\bar{1}0]$ (f), respectively.
  The red arrows show corresponding orientations of the induced electric 
  polarization.
}     
\end{figure}
We calculated the electric polarization induced in CuCrO$_2$ both by the
proper-screw as well as by cycloidal spin spirals with the propagation vectors
along the $\langle 1 {1} 0 \rangle$ and  $\langle 1 \bar{1} 0 \rangle$
directions. The results for CuCrO$_2$ are shown as a function of the
angle $\Theta = \vec{q}\cdot\vec{R}_{01}$ (with $\vec{R}_{01}$ the
position of a nearest-neighbor Cr atom) in
Figs.\ \ref{fig:CuCrO2-polar_screw} and \ref{fig:CuCrO2-polar_cycloid} 
for the proper screw and cycloidal spin modulations, respectively. 
  In the case of a proper-screw spin spiral with the propagation vector
  $\vec{q} || [110]$ ($\vec{q} || \hat{y}$),
  the electric polarization $\vec{P} || [110]$ is non-zero
with $\vec{P} \approx 30$~$\mu$C/m$^2$  for $\Theta = 120^o$, in
contrast to the prediction by the phenomenological KNB model
\cite{KNB05}. In the case of a
cycloidal spin spiral the non-vanishing in-plane polarization $\vec{P}
\approx 30$~$\mu$C/m$^2$ is found parallel to $[1\bar{1}0]$, in line with
experiment. These results indicate a dominating contribution due
  to the inverse-DMI (spin current) mechanism.
The remaining difference between the theoretical results and experiment
may be ascribed to a possible contribution due to other polarization
mechanisms \cite{TSN14}. 
For the proper screw spin spiral with $\vec{q} || [1\bar{1}0]$ ($\vec{q}
|| \hat{x}$) induced electric polarization is orthogonal to $\vec{q}$,
i.e. $\vec{P} || [110]$, similar to the results obtained for MnI$_2$.
In all cases, the electric polarization vanishes at $\Theta = 180^o$, when
the magnetic moments are getting collinear, changing sign due to a 
change of the spin spiral chirality. 
As one can see in Fig.\ \ref{fig:CuCrO2-polar_cycloid}(e) and (f), cycloidal
spin modulation induces also electric polarization along $z$ direction.
Occasionally, the polarization along the $z$ direction
almost vanishing at $\Theta = 120^o$ in the case of a cycloidal spin modulation
with the propagation vector $\vec{q} || [110]$ ($\vec{q} || \hat{y}$)
(see more details in Supplemental materials \cite{SuppMat}). 

The same trend can also be seen for AgCrO$_2$ \cite{SuppMat}.

\subsubsection {Cr$_2$O$_3$}

Finally, we consider the magnetoelectric properties of Cr$_2$O$_3$,
a prototypical material exhibiting the linear ME effect \cite{Dzy60,Ast61,KST79,BSD11}.
Cr$_2$O$_3$ crystallizes in the corundum structure with the
space group $R\bar{3}c$ (Nr. 167)  \cite{BFLT02}. Below $T_N = 310$ K it has an AFM structure with an alternating 'up-down-up-down'
($'udud'$) alignment of the Cr spin moments parallel to the hexagonal $c$
axis of the crystal (or alternatively, the rhomboidal $\langle 111 \rangle$ direction) \cite{Ast61}.
 Induced electric ($\vec{P}$) or magnetic ($\vec{M}$) polarization
  in the crystal appears as a linear response 
to an applied magnetic $\vec{H}_{\rm ext}$ or electric $\vec{E}_{\rm
  ext}$ field, respectively. The induced polarization is
described by the magnetoelectric tensor $\underline{\alpha}$ defined as
  $\alpha_{\mu\nu}^{H} = (\partial P_\mu/\partial H_{{\rm ext},\nu})$ in the
former and $\alpha_{\mu\nu}^{E} = \mu_0(\partial M_\mu/\partial E_{{\rm
    ext},\nu}$) in the latter case,  with only diagonal elements
being non-zero, $\alpha_{xx} = \alpha_{yy} = \alpha_{\perp}$ and
$\alpha_{zz} = \alpha_{||}$ (see, e.g. \cite{KST79,SBF+12,BLQ+24}).
On a microscopic level, the ME tensor can be obtained on the basis of the
DFT, calculating the response either to magnetic or to electric external
fields, respectively.
\begin{figure}[b]
  \includegraphics[width=0.13\textwidth,angle=0,clip]{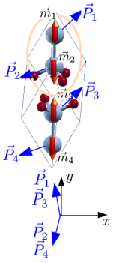}\,(a)
  \includegraphics[width=0.20\textwidth,angle=0,clip]{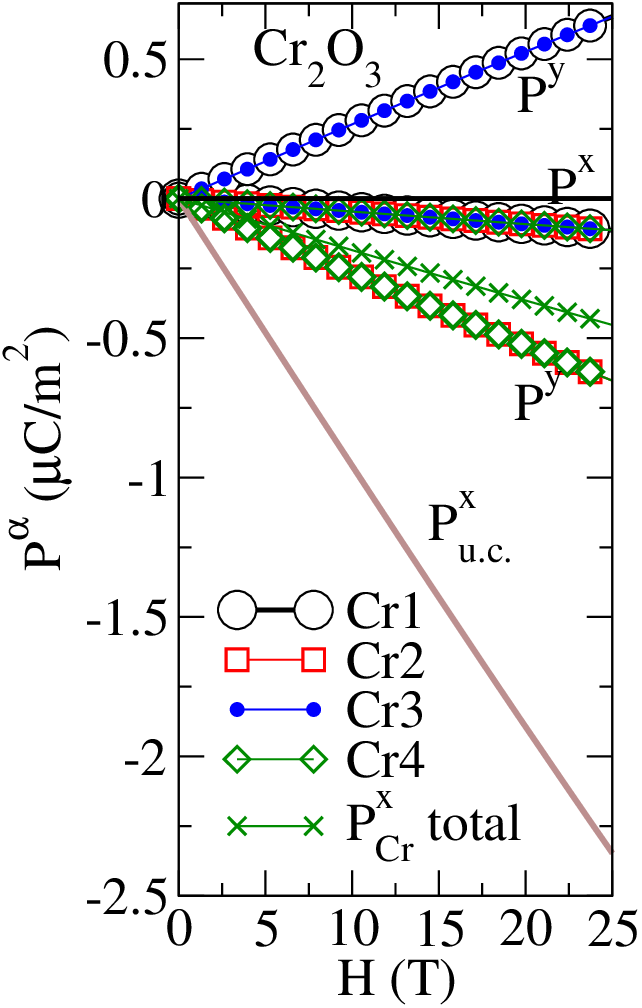}\,(b)\\
  \includegraphics[width=0.15\textwidth,angle=0,clip]{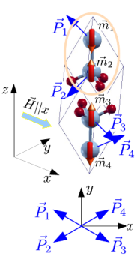}\,(c)
  \includegraphics[width=0.20\textwidth,angle=0,clip]{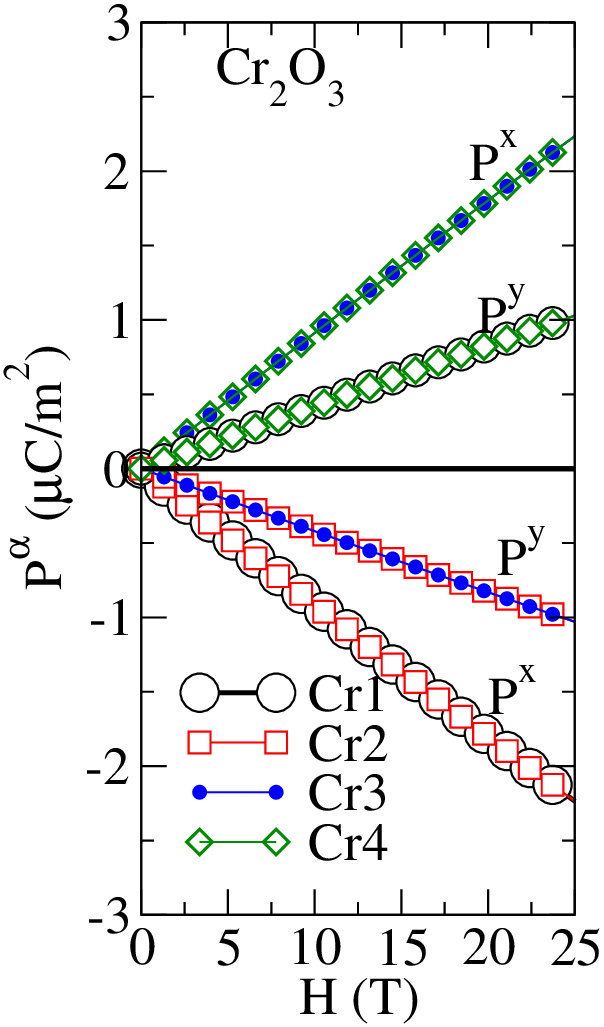}\,(d)
\caption{\label{fig_1:Cr2O3} Element-resolved induced electric
  polarization in Cr$_2$O$_3$ for the spin configurations $'udud'$
  (upper panel) and $'duud'$ (lower panel) (see text).
 Panels (b) and (d) show the  $P^x$ and $P^y$ components of the induced
 electric polarization of Cr atoms represented as a function of the tilting angle $\Theta$. Panels (a)
  and (c) show schematically the orientation of the polarization vectors in
  the unit cell (blue arrows). } 
\end{figure}

Here we discuss the ME effect of Cr$_2$O$_3$ in terms of the
parametrization scheme given in Section \ref{Section:theory}.
We focus on the contributions described by the parameters ${p}^{\gamma\nu,{\rm a}}_{ij,k}$, which are antisymmetric w.r.t.\ spin
permutation. The parameters are calculated making use of
Eq.\ (\ref{eq:dipol_KKR}), and then applied to the calculation of the
electric polarization induced by an external magnetic field via
Eq.\ (\ref{eq:dipol_6}).
The latter step implies spin rotations for the Cr atoms on sites $i$
by an angle $\Theta_i$ according to $({\rm sin}(\Theta_i), 0, {\rm cos}(\Theta_i))$, seen as a consequence
of an applied magnetic field $\vec{H}_{ext} = (H_x,0,0)$.
The electric polarization of the unit cell is obtained as a response to such spin
rotations, by summing over all induced electric polarizations
corresponding to 'magnetic'
(Cr) as well as to 'non-magnetic' (O) atoms, according to the
expression in the Eq.\ (\ref{eq:dipol_6}).   
When the magnetic field is assumed to be applied along the $x$ direction,
the non-zero contribution to the polarization stems from the parameters
${p}^{y\nu,{\rm a}}_{ij,k}$ (i.e. the component parallel to
  $\hat{s}_i \times \hat{s}_j$) with the index $\nu$ indicating 
the components of the polarization vector.

The components $P^x, P^y$ and $P^z$ of the induced electric polarization
of the sublattices associated to different Cr atoms are 
plotted as a function of the tilting angle $\Theta$ in Fig.\ 8 in the
Supplemental materials \cite{SuppMat}. The equilibrium angle $\Theta$ corresponding to
an imposed external magnetic field $H_{ext}$ was found using the
Heisenberg model with the calculated Cr spin moment $M = 2.5 \mu_B$ and
the Cr-Cr exchange coupling parameters \cite{SuppMat}.
Using these results, Figs.\ \ref{fig_1:Cr2O3} (b) and (d), represent the element projected
polarization components $P^x, P^y$ and $P^z$ as a function of the external
magnetic field $H_{ext}$.
Panels (a) and (c) show schematically the orientation of the
polarization vectors in the unit cell. Note that we consider here two
types of an AFM spin alignment in the unit cell, following 
the discussions in Ref.\ \onlinecite{Sol17}. 
Figs.\ \ref{fig_1:Cr2O3}, (a) and (b), show the results for the system with
the ground state spin configuration $'udud'$, which is invariant
w.r.t.\ the combined ${\cal IT}$ symmetry operation. One can see an identical
polarization of the Cr1 and Cr3 atoms which differ from the polarization
of the Cr2 and Cr4 atoms due to the opposite sign of their $P^y$
components. However, the sum over all Cr atoms in the unit cell is
finite and negative, while the $P^z$ component is equal to 0 for all
atoms.  Furthermore, the solid brown line in 
Fig.\ \ref{fig_1:Cr2O3}(b) represents the total polarization in the
unit cell accounting for the contributions from the O atoms, indicating its
strong enhancement due to the electric polarization of the O atoms.

The down-up-up-down ($'duud'$)
spin alignment (see Fig.\ \ref{fig_1:Cr2O3}(c),(d)) corresponds to the
system invariant w.r.t. the inversion operation ${\cal I}$, for which the electric
polarization should be forbidden by symmetry. This is seen in
Fig.\ \ref{fig_1:Cr2O3}(c) showing that the electric
polarization induced on different Cr atoms
are finite but cancel each other when summing up over the
unit cell, in line with the discussions by Solovyev \cite{Sol17}.
This is expected from symmetry considerations, as this spin
configuration ensures that the field-induced spin rotations
do not break inversion symmetry in the system.

Coming back to the ground state $'udud'$ spin configuration, one can
estimate the element $\alpha^H_{\perp} = (\partial
P_x/\partial H_{{\rm ext},x})$ of the ME tensor using the total
polarization per unit cell shown in Fig.\  \ref{fig_1:Cr2O3}(b). This leads to the
value $\alpha^H_{\perp} \approx - 0.12\, {\rm ps/m}$, which strongly
underestimates the experimental value that is in line with
other DFT based calculations demonstrating leading contribution of the
lattice mediated contribution to the total electric polarization
\cite{MCSV12a,BLQ+24}. 
Note that the estimation based on the polarization of the Cr sublattices only
gives $\alpha_{\perp} \approx -0.024$ ps/m, which is substantially
smaller and underestimates the available experimental data.

\section {Summary and outlook} \label{Section:summary}

In summary, we present in this work an approach to describe the ME
effect in insulating antiferromagnets on a first-principles
level.
This is of relevance first of all for materials with an extended non-collinear
magnetic structure, for which the developed real-space parametrization scheme
will in general be rather efficient, as it allows to avoid time-consuming supercell
calculations.
The approach presented may be seen as a generalization of the spin-current 
KNB model, giving a more complete description of the electronic
contribution to the induced electric polarization in real materials.
The results calculated for several representative materials are in full
agreement with other theoretical results obtained using different
generalization schemes for the KNB model.

A great benefit of using the KKR Green function formalism as a basis for
our approach is that it allows a number of extensions in various
directions. One is the treatment of disordered systems using the
Coherent Potential Approximation (CPA) alloy theory.
This can be exploited to study not only disordered alloys, but also the
temperature dependence of the ME parameters by using the alloy analogy
model (AAM) \cite{EMC+15}. Another important issue is the impact of
atomic displacements. This aspect was accounted for recently for the
exchange coupling constants leading to a scheme to calculate  the
spin-lattice coupling constants \cite{MPL+22,MLPE23}.
On that basis, another extension can be done for the approach presented
here allowing a description of 
the lattice-mediated polarization mechanism for the ME omitted so far. Work along these  directions is
in progress.

\section {Acknowledgements}

The authors gratefully acknowledge the computing time provided to them
on the high-performance computer Lichtenberg at the NHR Centers NHR4CES
at TU Darmstadt. 
J.M. was supported by the project Quantum materials for applications in
sustainable technologies (QM4ST), funded as project
No. CZ.02.01.01/00/22\_008/0004572 by P JAK, call Excellent Research.

\bibliography{multiferro}

\end{document}